\documentclass{jfm}

\usepackage{subfigure}
\usepackage{graphicx}
\usepackage{natbib}
\usepackage{bm}

\ifCUPmtlplainloaded \else
  \checkfont{eurm10}
  \iffontfound
    \IfFileExists{upmath.sty}
      {\typeout{^^JFound AMS Euler Roman fonts on the system,
                   using the 'upmath' package.^^J}%
       \usepackage{upmath}}
      {\typeout{^^JFound AMS Euler Roman fonts on the system, but you
                   dont seem to have the}%
       \typeout{'upmath' package installed. JFM.cls can take advantage
                 of these fonts,^^Jif you use 'upmath' package.^^J}%
      }
  \else
  \fi
\fi

\ifCUPmtlplainloaded \else
  \checkfont{msam10}
  \iffontfound
    \IfFileExists{amssymb.sty}
      {\typeout{^^JFound AMS Symbol fonts on the system, using the
                'amssymb' package.^^J}%
       \usepackage{amssymb}%
         \let\leq=\leqslant
         
      }{}
  \fi
\fi

\ifCUPmtlplainloaded \else
  \IfFileExists{amsbsy.sty}
    {\typeout{^^JFound the 'amsbsy' package on the system, using it.^^J}%
     \usepackage{amsbsy}}
    {}
\fi

\newsavebox{\astrutbox}
\sbox{\astrutbox}{\rule[-5pt]{0pt}{20pt}}

\title[Prandtl-Blasius temperature and velocity BL profiles in turbulent RBC]{Prandtl-Blasius temperature and velocity boundary layer profiles in turbulent Rayleigh-B\'{e}nard convection}

\author[Q. Zhou et al.]%
{Quan ZHOU$^1$, Richard J. A. M. STEVENS$^2$, Kazuyasu SUGIYAMA$^{2,3}$,
 Siegfried GROSSMANN$^4$, Detlef LOHSE$^2$,
\and
Ke-Qing XIA$^5$}

\affiliation{$^1$Shanghai Institute of Applied Mathematics and Mechanics, Shanghai University, Shanghai 200072, China \\
$^2$Physics of Fluids Group, Department of Science and Technology, J. M. Burgers Centre for Fluid Dynamics, and Impact-Institute, University of Twente, 7500 AE Enschede, The Netherlands \\
$^3$ Department of Mechanical Engineering, School of Engineering, The University of Tokyo, Tokyo, Japan\\
$^4$Fachbereich Physik, Philipps-Universit\"{a}t Marburg, D-35032 Marburg, Germany \\
$^5$Department of Physics, The Chinese University of Hong Kong, Shatin, Hong Kong, China}

\pubyear{1996}
\volume{538}
\pagerange{119--126}
\date{?? and in revised form ??}

\begin{document}

\maketitle

\begin{abstract}
The shape of velocity and temperature profiles near the horizontal
conducting plates in turbulent Rayleigh-B\'{e}nard convection are
studied numerically and experimentally over the Rayleigh number
range $10^8\lesssim Ra\lesssim3\times10^{11}$ and the Prandtl
number range $0.7\lesssim Pr\lesssim5.4$. The results show that
both the temperature and velocity profiles well agree
with the classical Prandtl-Blasius laminar
boundary-layer profiles, if they are re-sampled in the respective
dynamical reference frames that fluctuate with the instantaneous
thermal and velocity boundary-layer thicknesses.
\end{abstract}


\section{Introduction}

The turbulent motion in a fluid layer sandwiched by two parallel
plates and heated from below, i.e. Rayleigh-B\'{e}nard (RB)
convection, has become a fruitful paradigm for understanding the
physical nature of a wide range of complicated convection problems
occurring in nature and in engineering problems
(Siggia 1994; Ahlers, Lohse $\&$ Grossmann 2009; Lohse $\&$ Xia 2010). A key issue in the study of turbulent RB system is to understand how heat is
transported upwards by turbulent flow across the fluid layer. It is
measured in terms of the Nusselt number $Nu$, defined
as $Nu=J/(\kappa\Delta/H)$, which depends on the turbulent intensity and
the fluid properties. These are characterized, respectively, by
the Rayleigh number $Ra$ and the Prandtl number $Pr$, namely
$Ra=\alpha g H^{3} \Delta /\nu\kappa$ and $Pr=\nu/\kappa$. Here $J$
is the temperature current density across the fluid layer with a height
$H$ and with an applied temperature difference $\Delta$, $g$ the
gravitational acceleration, and $\alpha$, $\nu$, and
$\kappa$ are, respectively, the thermal expansion coefficient,
kinematic viscosity, and thermal diffusivity
of the convecting fluid, for which the Oberbeck-Boussinesq approximation is considered as valid.
As heat transport is controlled by viscosity and
thermal diffusion in the immediate vicinity of the solid boundaries, $Nu$ is
intimately related to the physics of the boundary layers.

In thermal convective turbulent flow two
types of boundary layers (BL) exist near the top and bottom plates, both of which are generated and
stabilized by the viscous shear of the large-scale mean flow:
One is the kinematic boundary layer and
the other is the thermal boundary layer. The two layers are not isolated
but are coupled dynamically to each other. They both play an
essential role in turbulent thermal convection, especially for the
global heat flux across the fluid layer. Almost
all theories proposed to predict the relation between $Nu$ and the control parameters
$Ra$ and $Pr$ are based on some kind of assumptions for the BLs, such
as the stability assumption of the thermal BL from the early
marginal stability theory \cite[]{malkus54}, the turbulent-BL
assumption from the theories of \cite{ss1990pra} and
\cite{sigga1994arfm} and of \cite{dubrulle2001ejpb,
dubrulle2002ejpb}, and the Prandtl-Blasius laminar-BL
assumption of the Grossmann $\&$ Lohse (GL) theory
\cite[]{gl2000jfm, gl2001prl, gl2002pre, gl2004pof}. Because of
the complicated nature of the problem, different theories based on
different assumptions for the BL may yield the same predictions
for the global quantities, such as the $Nu$-$Ra$ scaling relation
\cite[]{castaing1989jfm, ss1990pra}. Therefore, direct
characterization of the BL properties is essential for the
differences between and the testing of the various theoretical models and
will also provide insight into the physical nature of turbulent
heat transfer in RB system.

In the GL theory, the kinetic energy and thermal dissipation rates
have been decomposed into boundary layer and bulk contributions.
Scaling wise and in a time averaged sense a laminar Prandtl-Blasius boundary layer has
been assumed. This theory can successfully describe and predict
the Nusselt and the Reynolds number  dependences on $Ra$
and $Pr$ \cite[see e.g. the recent review in][]{agl2009rmp}. As
the Prandtl-Blasius laminar BL is a key ingredient of the GL theory, it is
important to make direct experimental verification of it.
We note that also the (experimentally verified)
calculation of the mean temperature in the bulk
in both liquid and gaseous non-Oberbeck-Boussinesq RB flows \cite[]{ahl06,ahl07,ahl08}
is based on the Prandtl-Blasius theory.

In a recent high-resolution measurement of the properties of the
velocity boundary layer, Sun, Cheung $\&$ Xia (2008) have found that,
despite the intermittent emission of plumes, the
Prandtl-Blasius-type laminar boundary layer description is indeed
a good approximation, in a time-averaged sense, both in terms of
its scaling and its various dynamical properties. However, because
of the intermittent emissions of thermal plumes from the BLs, the
detailed dynamics of both kinematic and thermal BLs in turbulent RB
flow are much more complicated. On the one hand, direct comparison
of experimental velocity (du Puits, Resagk $\&$ Thess 2007) and numerical
temperature \cite[]{thess2009jfm} profiles with theoretical
predictions has shown that both the classical Prandtl-Blasius
laminar BL profile and the empirical turbulent logarithmic profile
are not good approximations for the time-averaged velocity and
temperature profiles. Furthermore, \cite{sug09} from
two-dimensional (2D) and Stevens, Verzicco $\&$ Lohse (2010) from
three-dimensional (3D) numerical simulations found that the
deviation of the BL profile from the Prandtl-Blasius profile
increases from the plate's center towards the sidewalls, due to
the rising (falling) plumes near the sidewalls. On the other hand,
\cite{qiu1998prea} have found near the sidewall and
\cite{sun2008jfm} near the bottom plate that the velocity BL obeys
the scaling law of the Prandtl-Blasius laminar BL, i.e., its width
scales as $\lambda_v / H \sim Re^{-0.5}$, where $\lambda_v$ is the
kinematic BL thickness, defined as the distance from the wall at
which the extrapolation of the linear part of the local mean
horizontal velocity profile $u(z)=\langle u_x(z,t)\rangle$, with
$z$ being the vertical distance from the bottom plate and
$\langle\cdots\rangle$ being the time average at the plate center,
meets the horizontal line passing through the maximum horizontal
velocity $[u(z)]_{max}$, and $Re$ is the Reynolds number based on
$[u(z)]_{max}$. These papers highlight the need to study the nature
of the BL profiles, both velocity and temperature, in turbulent
thermal RB convection.

\begin{figure}
\begin{center}
\resizebox{0.61\columnwidth}{!}{%
  \includegraphics{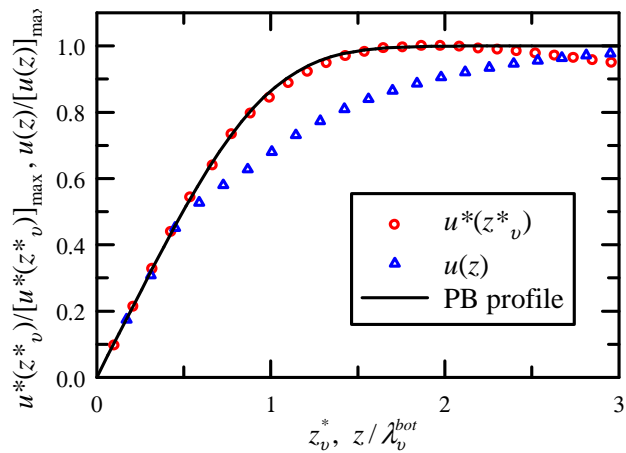}
}
\end{center}
\caption {Comparison between the spatial $x$-interval and time averaged velocity profiles
$u(z)$ (triangles), the dynamically rescaled velocity profile
$u^*(z^*_v)$ (circles -- for the notation we refer to section 3),
and the Prandtl-Blasius velocity profile (solid
line) near the bottom plate obtained experimentally at
$Ra=1.8\times10^{11}$ and $Pr=5.4$ (working fluid water).
}\label{fig1}
\end{figure}

Considerable progress on this issue has recently been achieved by
\cite{xia2009prl} who have experimentally studied the velocity BL
for water ($Pr=4.3$) with particle image  velocimetry (PIV). They
found that, since the dynamics above and below the range of the
boundary layer is different, a time-average at a fixed height $z$
above the plate with respect to the laboratory (or container)
frame will sample a mixed dynamics, one pertaining to the BL range
and the other one pertaining to the bulk, because the measurement
position will be sometimes inside and sometime outside of the
fluctuating width of the boundary layer. To make a clean
separation between the two types of dynamics, \cite{xia2009prl}
studied the BL quantities in a {\it time-dependent frame} that
fluctuates with the instantaneous BL thickness itself. Within this
dynamical frame, they found that the mean velocity profile well
agrees with the theoretical Prandtl-Blasius laminar BL profile. In
figure \ref{fig1} we show the essence of the results, again for
the velocity boundary layer but for somewhat larger $Pr$, now
$Pr=5.4$. (For details of the experiment and the apparatus used, please
see Xia, Sun $\&$ Zhou 2003; Zhou $\&$ Xia 2009). Also here the method of using the
time dependent frame works as good as for the $Pr=4.3$ case of
\cite{xia2009prl}. While at the large $Ra = 1.8 \times 10^{11}$
the time and space averaged velocity profile (triangles) already
considerably deviates from the Prandtl-Blasius profile (solid
line), the dynamically rescaled profile (circles) perfectly agrees
with the Prandtl-Blasius profile. Thus a dynamical algorithm has
been established to directly characterize the BL properties in
turbulent RB systems, which is mathematically well-defined and
requires no adjustable parameters.

The questions which immediately arise are: (i) Does this dynamical
rescaling method also work for the temperature field, giving good
agreement with the (Prandtl number dependent) Prandtl-Blasius
temperature profile? (ii) And does the method also work for lower
$Pr$, where the velocity field is more turbulent? Both these
questions cannot be answered with the current Hong Kong
experiments, as PIV only provides the velocity field and not the
temperature field, and as PIV has not yet been established in
gaseous RB, i.e., at low $Pr$ number Rayleigh-B\'{e}nard flows.

In the present paper we will
answer these two questions with the help of direct numerical simulations (DNS).
To avoid the complications of oscillations and rotations of the large scale convection roll plane
and as the Prandtl-Blasius theory is a 2D theory anyhow we will
restrict ourselves to the 2D simulations of \cite{sug09}.
Our results will show that \cite{xia2009prl}'s  idea of using
time-dependent coordinates to disentangle the mixed dynamics of BL and bulk works
excellently also for the temperature field and also for low $Pr$ flow.
I.e., if  dynamically rescaled, both velocity and temperature BL profiles can be
brought into excellent agreement with the theoretical Prandtl-Blasius
BL predictions, for both larger and lower $Pr$.

\section{DNS of the 2D Oberbeck-Boussinesq equations}
The numerical method has been explained in detail in \cite{sug09}.
In a nutshell, the Oberbeck-Boussinesq equations with no-slip velocity boundary conditions
at all four walls are solved for a 2D RB cell with a fourth-order finite-difference scheme.
The aspect ratio is $\Gamma \equiv D/L = 1.0$, the Rayleigh number
 $Ra= 10^8 - 10^9$, and the Prandtl number either $Pr=4.3$ (water) or
$Pr=0.7$ (gas). \cite{sug09} have provided a detailed code validation.

As the governing equations are strictly Oberbeck-Boussinesq, there
exists a top-bottom symmetry. We therefore discuss only the
velocity and temperature profiles near the bottom plate. For the
temperature profiles, we introduce the non-dimensional temperature
$\Theta(z,t)$, defined as
\begin{equation}
\Theta(z,t)=\frac{\theta^{bot}-\theta(z,t)}{\Delta/2},
\end{equation}
where $\theta^{bot}$ is the temperature of the bottom plate. In
this definition,  $\Theta(H)=2$ and $\Theta(0)=0$ are the
temperatures for the top and bottom plates, respectively, and
$\Theta(H/2)=1$ is the mean bulk temperature.

\section{Dynamical BL rescaling}

\begin{figure}
\begin{center}
\resizebox{1\columnwidth}{!}{%
  \includegraphics{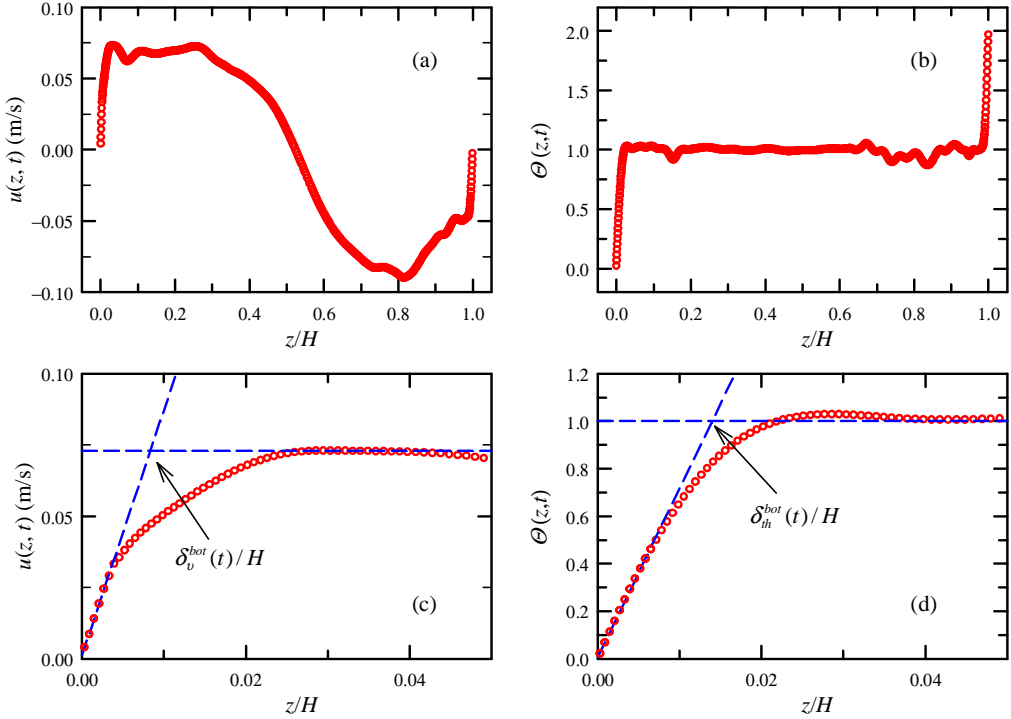}
}
\end{center}
\caption {Examples of (a) an instantaneous horizontal velocity
profile $u(z,t)$ and (b) a normalized instantaneous temperature
profile $\Theta(z,t)$, averaged over $0.475<x/D<0.525$. The DNS
data are obtained at $Ra=10^9$ and $Pr=0.7$. (c) and (d) show
enlarged portions of the velocity and temperature profiles near
the bottom plate, respectively. The two tilted dashed lines are
linear fits to the linear parts of the velocity and temperature
profiles near the plate and the two horizontal dashed lines mark
the instantaneous maximum horizontal velocity and the bulk
temperature $\Theta=1$, respectively. The distances of there crossing points
from the plate define the instantaneous BL thicknesses $\delta_{v,th}^{bot}(t)$.
The instantaneous profiles are not top-down symmetric, the time averaged ones are.
Within our present statistical error our data are consistent with zero thermal gradient in the bulk.
}\label{fig2}
\end{figure}

The idea of the \cite{xia2009prl} method is to construct a dynamical
frame that fluctuates with the local instantaneous BL thickness. To do
this, first the instantaneous kinematic and thermal BL thicknesses are
determined using the algorithm introduced by
\cite{xia2009prl}. To reduce data scatter, the horizontal velocity
and temperature profiles at each discrete time $t$, $u(z,t)$ and
$\Theta(z,t)$, are obtained by averaging the velocity and
temperature fields along the $x$-direction (horizontal) over the range
$0.475<x/D<0.525$. Figures \ref{fig2}(a) and
(b) show examples of $u(z,t)$ and $\Theta(z,t)$ versus the
normalized height $z/H$, respectively, of the DNS data obtained
at $Ra=10^9$ and $Pr=0.7$. Both $u(z,t)$ and
$\Theta(z,t)$ rise very quickly from 0 to either the instantaneous maximum velocity
or to the bulk temperature within very thin layers
above the bottom plate. While after reaching its maximum value,
$u(z,t)$ slowly decreases in the bulk region of the closed convection
cell, $\Theta(z,t)$ reaches and stays nearly constant at the bulk temperature
$\Theta=1$. To see the velocity and the
temperature in the vicinity of plates more resolved, we plot the
enlarged near-plate parts of the $u(z,t)$ and $\Theta(z,t)$
profiles in figures \ref{fig2} (c) and (d). One sees that both
profiles enjoy a linear portion near the plate. The
instantaneous velocity BL thickness $\delta_v(t)$ is then defined
as the distance from the plate at which the extrapolation of the
linear part of the velocity profile meets the horizontal line
passing through the instantaneous maximum horizontal velocity, and
the instantaneous thermal BL thickness $\delta_{th}(t)$ is
obtained as the distance from the plate at which the extrapolation
of the linear part of the temperature profile crosses the
horizontal line passing through the bulk temperature. The arrows
in figures \ref{fig2}(c) and (d) illustrate how to determine
$\delta_v(t)$ and $\delta_{th}(t)$ as the crossing point distances.

With these measured $\delta_v(t)$ and $\delta_{th}(t)$, we can now
construct the local dynamical BL frames at the plate's center.
The time-dependent rescaled distances $z^*_v(t)$ and $z^*_{th}(t)$ from the
plate in terms of $\delta_v(t)$ and $\delta_{th}(t)$, respectively, are defined as
\begin{equation}
z^*_v(t) \equiv z/\delta_v(t) \mbox{\ \ and\ \ } z^*_{th}(t) \equiv z/\delta_{th}(t).
\end{equation}
The dynamically time averaged mean velocity and temperature profiles $u^*(z^*_v)$ and
$\Theta^*(z^*_{th})$ in the dynamical BL frames are then obtained
by averaging over all values of $u(z,t)$ and $\Theta(z,t)$ that were
measured at different discrete times $t$ but at the same relative
positions $z^*_v$ and $z^*_{th}$, respectively, i.e.,
\begin{equation}
u^*(z^*_v) \equiv \langle u(z,t)|z = z^*_v\delta_v(t)\rangle \mbox{\ \
and\ \ } \Theta^*(z^*_{th}) \equiv \langle \Theta(z,t)|z = z^*_{th}\delta_{th}(t)\rangle.
\end{equation}

\begin{figure}
\begin{center}
\resizebox{1\columnwidth}{!}{%
  \includegraphics{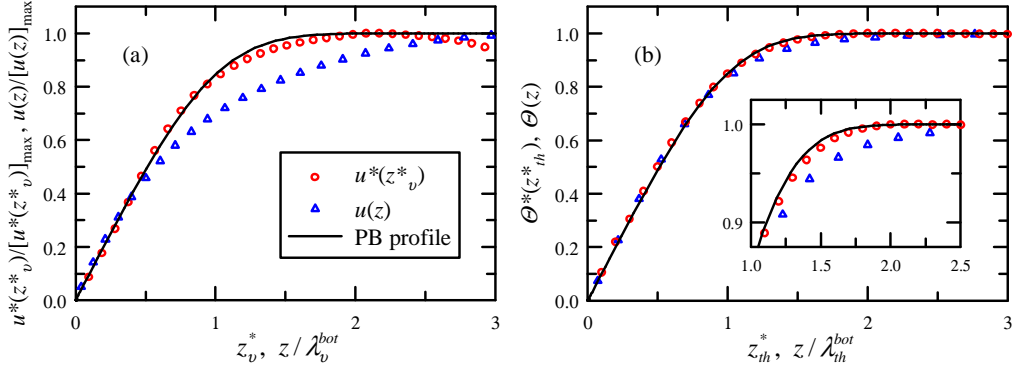}
}
\end{center}
\caption {Comparison among (a) velocity profiles: dynamical frame based $u^*(z^*_v)$
(circles), laboratory frame based $u(z)$ (triangles), and the Prandtl-Blasius velocity profile (solid
line), and (b) the corresponding temperature profiles: $\Theta^*(z^*_{th})$
(circles), $\Theta(z)$ (triangles), and the Prandtl-Blasius temperature
profile (solid line) near the bottom plate. All results obtained numerically at
$Ra=10^8$ and $Pr=4.3$. The inset of (b) shows enlarged portions
of the profiles around the thermal boundary layers' mergers to the bulk.
}\label{fig3}
\end{figure}

We first discuss our results from the simulation performed at
$Pr=4.3$, the Prandtl number corresponding to water at 40
$^{\circ}$C. Figure \ref{fig3}(a) shows the $u^*(z^*_v)$ profile
(circles), normalized by its maximum value $[u^*(z^*_v)]_{max}$,
obtained at $Ra=10^8$.  For comparison, we also plot in the figure
the time-averaged horizontal velocity profile $u(z)$ ($=\langle
u(z,t)\rangle$) (triangles), obtained from the same simulation.
The solid line represents the Prandtl-Blasius velocity BL profile, the initial
slope of which is matched to that of the measured profiles \cite[cf.][]{ahl06}.
For the range $z^*_v\lesssim2$ the $u^*(z^*_v)$ profile
obtained in the dynamical frame agrees well with the Prandtl-Blasius profile,
while the time-averaged $u(z)$ profile obtained in
the laboratory frame obviously is much lower than
the Prandtl-Blasius profile in the region around a few kinematic BL widths. - Note that for
$z^*_v\gtrsim2$ the $u^*(z^*_v)$ profile deviates gradually from
the Prandtl-Blasius profile because $u^*(z^*_v)$ decreases in the bulk region of
the closed convection system down to 0 in the center and then changes sign.
The Prandtl-Blasius profile, instead, describes the situation of an
asymptotically constant, nonzero flow velocity. -
These DNS results are similar to those found experimentally in a rectangular cell \cite[]{xia2009prl}.

Figure~\ref{fig3}(b) shows a direct comparison among the temperature
profiles obtained from the same simulation: the dynamical frame based $\Theta^*(z^*_{th})$
(circles), the laboratory frame time-averaged temperature profile $\Theta(z)$
($=\langle\Theta(z,t)\rangle$) (triangles), and the Prandtl-Blasius temperature
profile. At first glance both the $\Theta^*(z^*_{th})$ and
$\Theta(z)$ profiles are consistent with the Prandtl-Blasius thermal profile.
However, looking more carefully at the region around the thermal
BL to bulk merger (the inset of figure \ref{fig3}(b)), one notes that the
$\Theta^*(z^*_{th})$ profile obtained in the dynamical frame is significantly
closer to the Prandtl-Blasius profile than the time-averaged $\Theta(z)$
profile obtained in the laboratory frame, indicating that the
dynamical frame idea of \cite{xia2009prl} works also for the thermal BL.
Taken together, figures \ref{fig3}(a) and (b) illustrate that both the
kinematic and the thermal BLs in turbulent RB convection are of Prandtl-Blasius type,
which is a key assumption of the GL theory \cite[]{gl2000jfm,
gl2001prl, gl2002pre, gl2004pof}, and the dynamical frame idea of
\cite{xia2009prl} can achieve a clean separation for both
temperature and velocity fields between their BL and bulk dynamics.

\begin{figure}
\begin{center}
\resizebox{1\columnwidth}{!}{%
  \includegraphics{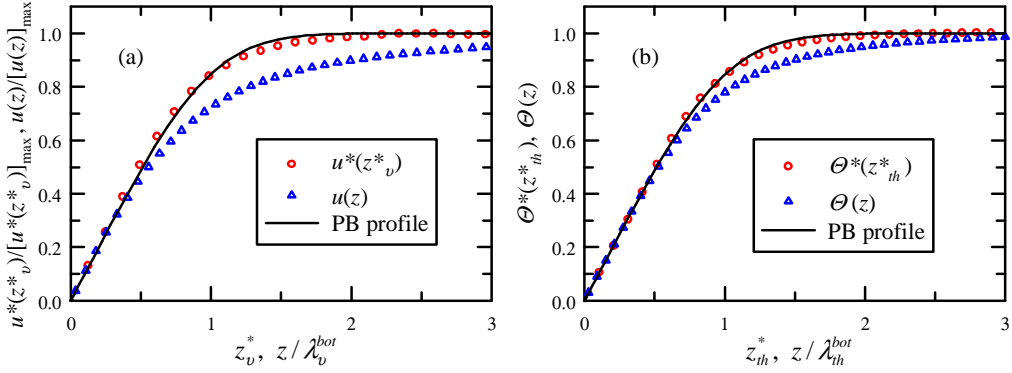}
}
\end{center}
\caption {Comparison between (a) velocity profiles: dynamical $u^*(z^*_v)$
(circles), laboratory $u(z)$ (triangles), and the Prandtl-Blasius laminar velocity profile (solid
line), and (b) temperature profiles: dynamical $\Theta^*(z^*_{th})$
(circles), laboratory $\Theta(z)$ (triangles), and the Prandtl-Blasius laminar temperature
profile (solid line) near the bottom plate, all obtained numerically at
$Ra=10^9$ and $Pr=0.7$, representative for gases.
}\label{fig4}
\end{figure}

We next turn to the simulation performed at $Pr=0.7$, a Prandtl number
typical for gases, which is relevant in all atmospheric
processes and many technical applications. Figures \ref{fig4}(a)
and (b) show direct comparison between the temperature and velocity
profiles, respectively, at $Ra=10^9$. Again, around the BL-bulk merger range the laboratory frame
time-averaged profiles are found to be obviously lower than
the Prandtl-Blasius profile. This once more indicates that the time-averaged BL
quantities obtained in the laboratory frame are contaminated by
the mixed dynamics inside and outside the fluctuating BLs. On the
other hand, within the dynamical frame, both $u^*(z^*_v)$ and
$\Theta^*(z^*_{th})$ are found to agree pretty well with the Prandtl-Blasius laminar BL
profiles, indicating that the dynamical frame idea works also for the
turbulent RB system with working fluids whose Prandtl numbers are
of the same order as those for gases.

\section{Shape factors of the velocity and temperature profiles}

Let us now quantitatively compare the differences between the
Prandtl-Blasius profile and the profiles obtained from both simulations and
experiments for various $Ra$ and various $Pr$. The shapes of the
velocity and temperature (thermal) profiles, labeled by $i = v$ or $i = th$, can be
characterized quantitatively by their shape factors $H_i$, defined as
\cite[]{schlichting04},
\begin{equation}
H_i=\frac{\lambda^d_i}{\lambda^m_i} ~, ~~~ i = v,th.
\end{equation}
$\lambda^d_i$ and $\lambda^m_i$ denote, respectively, the
displacement and the momentum thicknesses of the profile, namely,
\begin{equation}
\label{eq. Shape}
\lambda^d_i=\int_0^{\infty}\{1-\frac{Y(z)}{[Y(z)]_{max}}\}dz
\mbox{\ \ and\ \ }
\lambda^m_i=\int_0^{\infty}\{1-\frac{Y(z)}{[Y(z)]_{max}}\}\{\frac{Y(z)}{[Y(z)]_{max}}\}dz.
\end{equation}
Here $Y(z)=u(z)$ is the velocity profile if $i=v$ and $Y(z)=\Theta(z)$ the thermal
profile if $i=th$. The deviation of these profiles from the Prandtl-Blasius profile is then measured by
\begin{equation}
\delta H_i =  H_i - H^{PB}_i ,
\end{equation}
where $H^{PB}_i$ is the shape factor for the respective
Prandtl-Blasius laminar BL  profile. If a given profile exactly
matches the Prandtl-Blasius profile, $\delta H_i$ is zero. Note
that the Prandtl-Blasius velocity profile shape factor
$H^{PB}_v=2.59$ is independent of $Pr$, while the thermal
Prandtl-Blasius BL profile shape factor $H^{PB}_{th}$ varies with
$Pr$.

\begin{figure}
\begin{center}
\resizebox{1\columnwidth}{!}{%
  \includegraphics{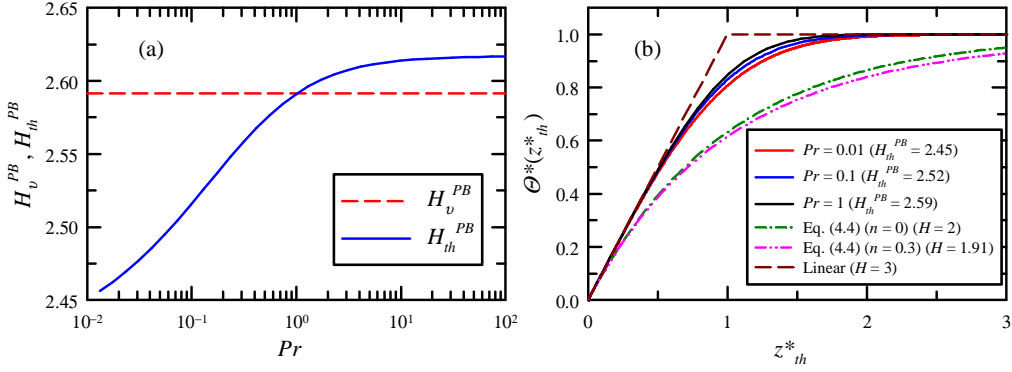}
}
\end{center}
\caption {(a) The shape factors for the thermal (solid line) and
velocity  (dashed line) Prandtl-Blasius BL profiles as function of
$Pr$. The asymptotic value $H^{PB}_{th}(Pr \gg 1)$ is $2.61676...$
 and $H^{PB}_v
= 2.59$. Both Prandtl-Blasius BL profiles for the velocity and for
the temperature coincide for $Pr=1$. (b) The thermal
Prandtl-Blasius BL profiles for three (four) $Pr$ numbers and the
reference linear and exponential profiles, see (\ref{eq. exp}); in
the figure's resolution the thermal profile for $Pr=100$ is
indistinguishable from that of $Pr=1$.  Note that the shape factor
of the thermal Prandtl-Blasius BL profile decreases with
decreasing $Pr$ due to the slower approach to its asymptotic level
$1$. } \label{fig5}
\end{figure}

\begin{figure}
\begin{center}
\resizebox{1\columnwidth}{!}{%
  \includegraphics{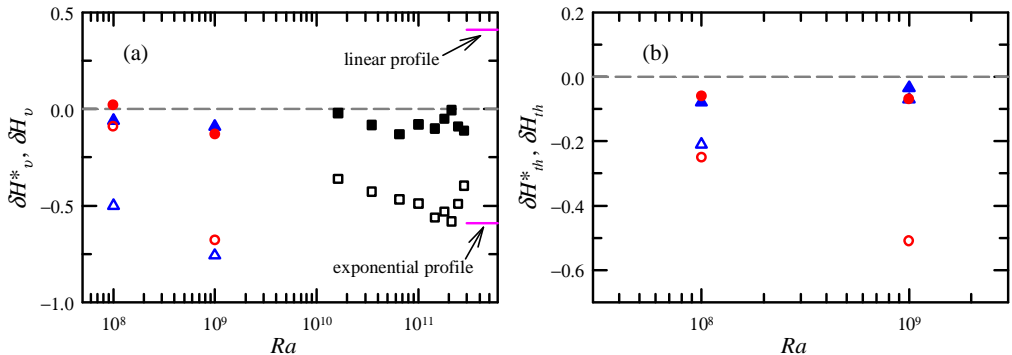}
}
\end{center}
\caption {The $Ra$-dependence of the deviations of the profile
shape factors from the respective Prandtl-Blasius shape factors. (a) Laboratory frame $\delta H_v$ (open symbols)
and dynamical frame $\delta H^*_v$  (solid symbols); (b) laboratory frame $\delta H_{th}$ (open
symbols) and dynamical frame $\delta H^*_{th}$ (solid symbols); all from simulations
performed at $Pr=0.7$ (circles), $Pr= 4.3$ (triangles), and from experiments at $Pr=5.4$ (squares).
}
\label{fig6}
\end{figure}

Figure \ref{fig5}(a) shows the shape factors $H_i(Pr)$ of the
thermal and the velocity  Prandtl-Blasius BL profiles as functions
of $Pr$ and figure \ref{fig5}(b) shows the corresponding thermal
profiles as functions of $z^*_{th}$ for three different $Pr$. Note
that the Prandtl-Blasius velocity BL profile is identical to the
thermal one for $Pr=1$. The two figures show that the thermal
shape factor $H_{th}^{PB}$ decreases with decreasing $Pr$. We attribute this to
the decrease of the temperature profiles in the BL range and the corresponding
increase of the tails for lower $Pr$. Thus we expect that the
slower approach to the asymptotic height $1$ of the thermal
profiles in the laboratory frame in figures \ref{fig3} and
\ref{fig4} should lead to a negative deviation of their $H_{th}$'s
from the respective Prandtl-Blasius values, cf. figure \ref{fig6}.
In contrast, a positive $\delta H_i$ is obtained if the profile
runs to its asymptotic level faster than the Prandtl-Blasius
profile. To see this more clearly, we have plotted in figure
\ref{fig5}(b) also two extreme cases, the linear and the
exponential profiles. Using (\ref{eq. Shape}) one calculates  the
shape factor $3$ for the linear profile
$\Theta^*(z^*_{th})=\mbox{min}(1,z^*_{th})$ and the shape factor
$2$ for the exponential one
$\Theta^*(z^*_{th})=1-\mbox{exp}(-z^*_{th})$. The $H$-decreasing
effect by lowering the profile can also be demonstrated by
analyzing some profiles analytically. Using a combination of
exponential profiles,
\begin{equation}\label{eq. exp}
 \Theta^*(z^*_{th})=0.5(1-\mbox{exp}(-(1-n)(z^*_{th}))+0.5(1-\mbox{exp}(-(1+n)z^*_{th})),
\end{equation}
  with $0\leq n < 1$ one can evaluate, using (\ref{eq. Shape}), that the shape factor for small $n$ is
\begin{equation}\label{}
    H(n) \approx  H(n=0) -n^2 = 2 -n^2.
\end{equation}
As is shown in figure \ref{fig5}(b) the profile for $n>0$ is below the profile for $n=0$. This analytical example again reflects what we found as the characteristic difference between the laboratory frame profiles as compared to the dynamical frame profiles.

Figure \ref{fig6}(a) shows the velocity shape factor deviations $\delta H_v$ (open symbols)
and $\delta H^*_v$ (solid symbols) as obtained from simulations at $Pr=0.7$ (circles) and
$Pr = 4.3$ (triangles) as well as from experiments at $Pr=5.4$ (squares).
Here, $\delta H_v$ is calculated with the time averaged profile $u(z)$ in the
laboratory frame, while $\delta H^*_v$ is calculated with the dynamical, time-dependent frame profile $u^*(z^*_v)$.
The laboratoty frame based deviations turn out to be definitely smaller than zero. In contrast,
the shape factor deviations $\delta H^*_v$ for the dynamical frame profiles obviously are much
closer to zero.  A similar result is found for the thermal BLs: Figure \ref{fig6}(b) shows $\delta H_{th}$
(open symbols) and $\delta H^*_{th}$ (solid symbols), versus $Ra$,
for the same $Pr$ number simulations. Again $\delta H^*_{th}$ is
nearly zero, whereas $\delta H_{th}$ is significantly off. Thus these quantitative deviation measures
again indicate that the algorithm using the dynamical coordinates can effectively disentangle
the mixed dynamics inside and outside the fluctuating BLs.

\section{Conclusions}

In summary, we have studied the velocity and temperature  BL
profiles in turbulent RB convection both numerically and
experimentally. We extended previous results to different Prandtl
numbers and in particular to thermal BLs. The results show that both the velocity
and the temperature BLs (at least in the plates' center region) are of laminar
Prandtl-Blasius type in the co-moving dynamical frame in turbulent thermal convection for
the parameter ranges studied. However, the fluctuations of the BL widths,
induced by the fluctuations of the large-scale mean flow and the
emissions of thermal plumes, cause measuring probes at fixed heights above
the plate to sample a mixed dynamics, one pertaining to the BL range and the other one
pertaining to the bulk. This is the reason why the time-averaged
velocity and temperature profiles measured in previous work in fixed
laboratory (RB cell) frames deviate  from the Prandtl-Blasius profiles. To
disentangle that mixed dynamics, we constructed a dynamical BL frame
that fluctuates with the instantaneous BL thicknesses. Within this dynamical
frame, both velocity and temperature profiles are very well consistent with the
classical Prandtl-Blasius laminar BL profiles, both for lower and larger $Pr$ (from $0.7$ to $5.4$).
We have thus validated the idea and algorithm of using dynamical coordinates
over a range of $Pr$ and $Ra$ for both kinematic and thermal BLs
and have shown that the Prandtl-Blasius laminar BL profile is a valid description for
the BLs of both velocity and temperature in turbulent thermal convection. Laminar
Prandtl-Blasius BL theory in turbulent RB thermal convection has thus turned out
to indeed be valid not only scaling wise, but also in the time average as seen
from the dynamical frame, co-moving with the local, instantaneous BL widths.

\begin{acknowledgments}
We gratefully acknowledge support of this work by the
Natural Science Foundation of Shanghai (No. 09ZR1411200),
``Chen Guang" project (No. 09CG41)(Q.Z.), by the Research
Grants Council of Hong Kong SAR (Nos. CUHK403806 and 403807) (K.Q.
X), and by the research programme of FOM, which
is financially supported by NWO (R.J.A.M.S. and D.L.).
\end{acknowledgments}

%
%

\bibliographystyle{jfm}

\end{document}